*See NOTE below, at end of this file, if you encounter problems with hyperlinks cited in text and listed in "Hyperlinks Cited."  This front-end statement and the end-of-file addition are the **only** changes here, from 5Apr18 version of document.*

## Consilience:  A Holistic Measure of Goodness-of-Fit

William H. Neill[1], Ray H. Kamps[2], Scott J. Walker[3], Hsin-i Wu[4], T. Scott Brandes[5], Delbert M. Gatlin III[1], Tiffany L. Hopper[6], and Robert R. Vega[7]

[1] Department of Wildlife & Fisheries Sciences, Texas A&M University System, 2258 TAMUS, College Station, Texas 77843-2258

[2] Department of Soil & Crop Sciences, Texas A&M University System, 2474 TAMUS,  College Station, Texas 77843-2474

[3] U.S. Fish and Wildlife Service, Inks Dam NFH, 345 Clay Young Road, Burnet, TX  78611

[4] Department of Biomedical Engineering, Texas A&M University, College Station, TX 77843-3120

[5] BAE Systems, Inc., 4721 Emperor Blvd., Suite 330, Durham, NC  27703

[6] Texas Parks & Wildlife Department, Coastal Fisheries Division, 4200 Smith School Road, Austin, TX 78744

[7] Texas Parks & Wildlife Department, CCA Marine Development Center, Coastal Fisheries Division, Corpus Christi, TX  78418

## Abstract

We describe an apparently new measure of multivariate goodness-of-fit between sets of quantitative results from a model (simulation, analytical, or multiple regression), paired with those observed under corresponding conditions from the system being modeled.  Our approach returns a single, integrative measure, even though it can accommodate complex systems that produce responses of M types.  For each response-type, the goodness-of-fit measure, which we label "Consilience" (C), is maximally 1, for perfect fit; ~0 for the large-sample case (number of pairs, N, > ~25) in which the modeled series is a random sample from a quasi-normal distribution with the same mean and variance as that of the observed series (null model); and, < 0, toward -∞, for progressively worse fit. In addition, lack-of-fit for each response-type can be apportioned between systematic and non-systematic (unexplained) components of error.  Finally, for statistical assessment of models relative to the equivalent null model, we offer provisional estimates of critical C vs. N, and of critical joint-C  vs. N and M, at various levels of Pr(type-I error).  Application of our proposed methodology requires only MS Excel (2003 or later); we provide Excel XLS and XLSX templates that afford semi-automatic computation for systems involving up to $M = 5$ response types, each represented by up to $N = 1000$ observed-and-modeled result pairs.  N need not be equal, nor response pairs in complete overlap, over M.

**Acknowledgements**

We thank F. Michael Speed for his early encouragement, guidance, and practical assistance with development of ideas and methodology for assessing holistic goodness-of-fit.

We are grateful also for the financial support provided by the Coastal Fisheries Division of Texas Parks and Wildlife Department, and by Texas A&M University System.

**Contents**

**Introduction**

Probably, most of us rely on the familiar "coefficient of determination"—computed as the squared correlation coefficient ($r^2$ from simple regression; or, $R^2$ from multiple regression[1])—to gauge how well our models explain the variation in our data. But, such "explanation" may fail utterly, when the goal is to measure how well modeled outcomes agree with observed outcomes, or predict outcomes yet to be observed.

As a good example of the problem, consider the flawed performance of the nominal version of the simulation model Ecophys.Fish (Neill et al. 2004), in simulating observed growth rates of the bluegill (a centrarchid sunfish)—this, despite an $R^2$ of 0.83 for 10 independent pairs of values; an "optimized" variant of the model clearly provided a better fit to the data, albeit with a decline in $R^2$ to 0.77 (Fig. 1; redrawn here from Fig. 22, Neill et al. 2004). The underlying problem is that the coefficient of determination essentially measures goodness-of-fit to the regression line, not to the line of perfect agreement between modeled and observed values.

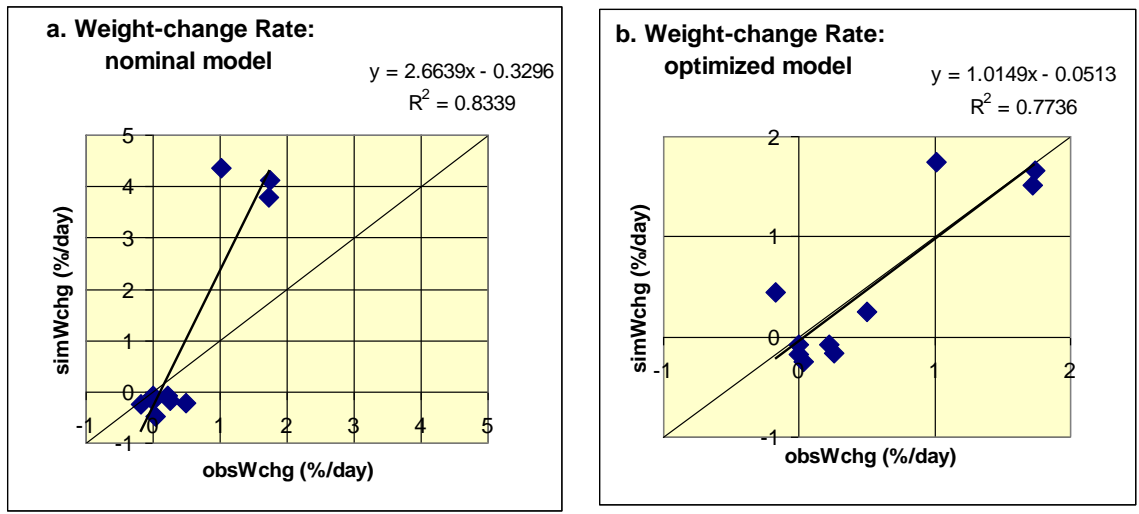

*Figure 1. Example demonstrating that the better-fitting model may not present the higher value of $R^2$. (Redrawn from Neill et al. 2004, Fig. 22: Observed vs. simulated rates of weight change, in field test of nominal model (a), and under the optimized model (b), for bluegill.)*

More directed approaches to assessment of goodness-of-fit (GoF) have a long history. For categorical data, there is the familiar Pearson's chi-square (Pearson 1900) and its successors (e.g., G-test; Sokal and Rohlf 1981, 2011); and, Fisher's exact-probability test (Fisher 1935), and its multinomial extensions.

---

[1] Throughout this paper we use $R^2$, in lieu of $r^2$, because we consider analysis of modeled vs. observed responses of natural systems to be more akin to multiple regression than to simple regression.

These are capably treated, both in-print (e.g., in conventional textbooks, such as Snedecor and Cochran 1989; Freund and Wilson 2003; and, Heiberger and Holland 2004), and on-line (e.g., John McDonald's Handbook of Biological Statistics).  But, these goodness-of-fit tests amount to protocols for rejecting bad models, statistically—not for measuring performance of models intended to capture and portray mechanistic truth about the system being modeled.  Moreover, their focus is on differences between frequency distributions of observed vs. modeled values, not on pair-wise comparison.  In the context of pair-wise GoF, Schunn and Wallach (2005) provide a more general exploration and discussion of analytical approaches; we call attention, especially, to the Excel template offered via their footnote 5, on their p. 130.

When we considered the problem of comparing, quantitatively, models and observed systems that present N presumably unique continuous-response pairs for each of 1 to M response types, we found no existing methodology for measuring what we have come to regard as holistic goodness-of-fit (HGoF).  So, we decided to try developing a methodology of our own.  We sought a simple, transparent approach that could be implemented via a tool-set no more specialized than Excel.  Moreover, we wanted a methodology that could sensibly partition lack-of-fit, between sources of error that are systematic vs. those that remain ("noise").  Finally, we wanted to reasonably integrate the multivariate components of goodness-of-fit, to arrive at a single number, one that scaled from 1, for perfect fit, then downward for increasingly bad fit.

This single number at which we eventually arrived, and that we offer as a holistic measure of goodness-of-fit, we have labeled "Consilience," with "C" for the acronym.  The word "consilience," made popular by E.O. Wilson's book *Consilience: The Unity of Knowledge* (Wilson 1998), traces back at least as far as 19th-century theologian and Cambridge philosopher-of-science William Whewell, who used the word to mean an effective union of facts and ideas, especially those from disparate sources (Whewell 1840; but, see also "William Whewell" by Snyder 2012).  Our choice of the label "Consilience" is intended to pay homage to the Rev. Whewell and to Dr. Wilson, for their insights regarding systemic understanding of reality.

**Toward Quantification of "Consilience," C**

Normally, we modelers build models to generate values of a single, focal output of the observed system—no matter how many are the inputs nor how complex the network of presumed cause-and-effect between input(s) and output.  But, simulation models can (and, we think *should*) be designed to produce responses of multiple types, to represent the multiple and interrelated responses of the modeled system.

For example, our model Ecophys.Fish simulates metabolic responses, bioenergetics and growth of fish, relative to time-varying, multivariate environment. Its instructional successor, [EcoFish](EcoFish), extends that suite of responses to include others as diverse as stomach evacuation, resistance to lethal factors and enviroregulatory behavior.  These responses in the real fish no doubt are physiologically interrelated, and so too must they be represented in any model that presumes to be mechanistic.  How can the M types of responses, represented by N cases, be properly quantified and weighted so that the model can be evaluated as a whole?

*Resolution of Modeling Error and Computation of C*

The search for a general answer to this question began with the obvious and simple idea, that the data observed from a system represent our best evidence of that system's true workings, at least on that particular occasion (or set of occasions).  Certainly, two or more observations on the system may present different measured values of some output despite apparently identical inputs; that, we can attribute to measurement error, or to random error (stochasticity)—at the same time, realizing full well that the "noise" might have been caused by unmeasured input variables that differed between or among instances.  Conversely, differing sets of inputs may yield the same output, implying compensatory errors or that the response is not monotonic.  But, in any event, the ultimate goal of a model logically must be to yield values that are identical with corresponding observations from the modeled system, in every instance.

Starting with that goal in mind, modeling error then must be rendered as

TotalError = Yobs - Ymod,

where Yobs and Ymod are the observed and modeled values for each response pair.  Now, partition TotalError into systematic and non-systematic parts, by subtracting and adding Yp, the value of Y from linear regression of Ymod on Yobs:

TotalError = Yobs - Yp   +  Yp - Ymod

$\qquad$ = (Yobs - Yp) + (Yp - Ymod)

Logically, the first right-hand term in parentheses is the systematic error in modeled values; whereas, the second is the non-systematic error in those values (or "random error," or "noise").  Together, the two must represent total error.

TotalError = (Yobs - Yp)   +   (Yp - Ymod)

$\qquad$ =    SysErr     +    RanErr .

Figure 2 is intended to illustrate, in a graphical way, the relationships among Yobs, Ymod, Yp, TotalError and its two components.

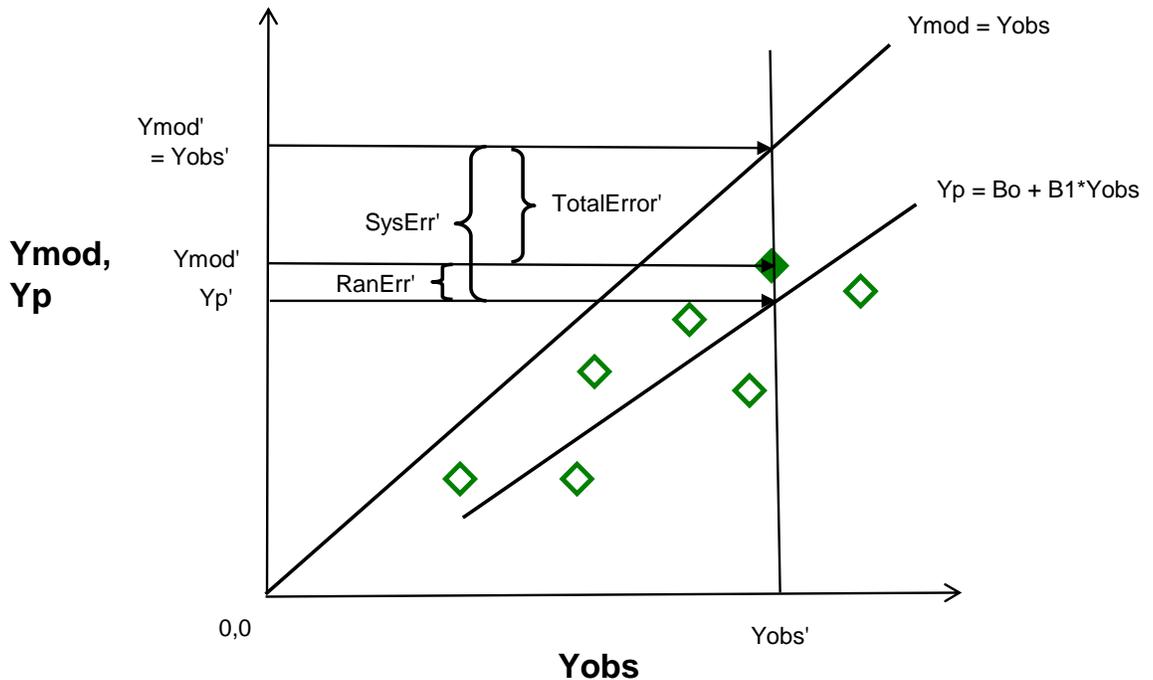

*Figure 2. Graphical illustration of conceptual relationships among Yobs, Ymod, Yp, TotalError, and its components SysErr (systematic error) and RanErr (non-systematic error, or "noise"). Note that*

$$TotalError = SysErr + RanErr ;$$

$$Yobs - Ymod = (Yobs - Yp) + (Yp - Ymod) .$$

In Figure 2, the particular value of Ymod,Yobs was selected to make the point that components of TotalError can be either positive or negative. In this specific instance, SysErr' = (Yobs' - Yp') is positive, but RanErr' = (Yp' - Ymod') is negative. Because Yp is the result of least-squares regression of Ymod on Yobs, the mean of residual errors Yp - Ymod always will be zero; however, for unconstrained simulation models, the mean of systematic errors Yobs - Yp normally will be non-zero.

Anticipating the need to arrive eventually at squared-error terms motivated rearrangement of TotalError components, as a difference of differences:

TotalError = (Yobs - Yp) - (Ymod - Yp) .

Next, we considered that various datasets most likely will present differing distributions of Yobs, both for differing responses of the same system, and for the same response of different systems. Therefore, one might scale for some measure of variation in the distributions of observed response, by dividing all terms by that common measure:

TotalError =  (Yobs - Yp) -  (Ymod - Yp)
   scalar        scalar           scalar

Two alternative scalars, to which we have given main consideration, are the standard deviation of Yobs, and the interquartile range of Yobs. For reasons largely heuristic (see below), we have come to prefer as scalar the standard deviation over the interquartile range—although the associated templates accommodate use of either (and of the mean or median of Yobs, as well; see below, regarding templates).

Next, squared scaled-error terms can be computed:

$$(TotalError/scalar)^2 = ((Yobs - Yp)/scalar)^2 + ((Ymod - Yp)/scalar)^2$$

$$- 2*((Yobs-Yp)/scalar)*((Ymod - Yp)/scalar).$$

Restating the last equation,

ScaledTotErrSqrd    =    ScaledSysErrSqrd   +   Scaled"Ran"ErrSqrd
                            - 2*ScaledSysErr*Scaled"Ran"Err .

We initially reasoned that if systematic and "random" components of TotalError are uncorrelated, then the expected value of the cross-product term is zero, leaving the expected value of ScaledTotErrSqrd equal to the sum of expected values of its systematic and non-systematic parts.

But, it turns out that not only is the expected value of the cross-product term zero, but also—in every case—the computed mean cross-product is, in fact, zero, and thus vanishes (provided that the scalar is constant among terms and over all Yobs within the sample). See Appendix I for algebraic support of this claim—or, if you would prefer, go to one of the templates (HGoFtemplate.xls or ...xlsx; see below) and run hypothetical cases until you are convinced.

We conclude that

Mean(ScaledTotErrSqrd)

        = Mean(ScaledSysErrSqrd) + Mean(Scaled"Ran"ErrSqrd);   or,

MSEtot = MSEsys + MSEran.

Thus, the proportions of Mean(ScaledTotErrSqrd) = MSEtot attributable to systematic and non-systematic components can be computed simply as the ratios,

MSEsys/MSEtot and MSEran/MSEtot, respectively.

It is apparent, in the case of perfect model-fit, that MSEtot, MSEsys, and MSEran all will be zero. As the model's fit deteriorates, the values of the mean-squared errors inflate. Because we wanted a measure of goodness-of-fit that scaled, in the perfect-fit limit, like the coefficient of determination from regression ($r^2$ or $R^2$), we decided to focus on MSEtot and transform it as

$$-(MSEtot - 2)/2$$

and call the transform "C," for Consilience:

$$\boxed{C = -(MSEtot - 2)/2.}$$

Why subtract 2, then divide by 2 (as opposed to, say, 1 and 1)? The choice arose as the consequence of a historical happenstance, to keep C within the bounds [-1,1], when we first tried an approach based on standardized residuals (see HGoFpres1.avi or ...mov). Having abandoned that approach as inadequate (see HGoFpres2.avi or ...mov) and moved on to the one presented here, we decided to leave the transform as it was, when we noticed that using 2 and 2 not only produces C = 1 in the event of perfect model-fit, but also that the 2-and-2 choice makes mean C approach 0 as $N \rightarrow \infty$, provided Ymod values are randomly sampled from a quasi-normal distribution with the same mean and variance as Yobs, and provided that the scalar is the common standard deviation. We thought that constituted a convenient point of reference and sound basis for scalar choice.

*Introduction of Holistic Goodness-of-Fit (HGoF) Template and the Compact C-calculator*

A generic Excel template for semi-automatic computation of squared-scaled-error terms and C, for systems involving M up to 5 and N up to 1000, is provided by HGoFtemplate.xls (or ...xlsx). The template repeats our conceptual derivation of HGoF; executes HGoF in explicit detail, for a dataset inserted by the user; estimates probability of larger values of C, by chance; and, for comparison with computed C, offers graphs for visualization and conventional statistical procedures for assessing goodness-of-fit. The template opens to a sheet with essential "User Guidance" (not all repeated in the text of this paper), taking the user through HGoF, step-by-step.

Once comfortable with the ideas and template, the user may want to work instead with our CompactCCalculator.xlsx. This very small analog of the HGoFtemplate uses matrix-based operations to duplicate the essential steps of HGoF, and generates identical key outputs for the same inputs. Not only is CompactCCalculator less than 100Kb in size, but also it accommodates datasets with indefinitely large N and allows the user to re-order response components (for M up to 5), electronically. Like HGoFtemplate, CompactCCalculator can process datasets with uneven N over M, and with missing data. However, we are unable to provide an XLS version of the CompactCCalculator.

For would-be users without access to Excel, we encourage opening and trying the template and/or calculator with OpenOffice Calc for Windows, or LibreOffice Calc for Linux. Examine your outputs to be sure they are consistent with summary results provided in ExampleDataSets.xls (or ...xlsx) for the same input arrays. Expect some display flaws that, hopefully, are minor and only cosmetic.

The example applications described in the next section and the one following, below, involve only systems for which M = 1. We defer consideration of M > 1 systems accommodated by the template and the compact calculator, until the section, "Joint C: Weighting of Multiple Responses for HGoF," below.

*An Example Application*

As a first example (and recommended exercise for the user), consider the model-vs.-data comparisons shown in Figure 1. The requisite inputs are those appearing in datasets WchgBGnom and WchgBGopt of sheet M=1, ExampleDataSets.xls (and ...xlsx). The computed values of C are -1.18 for the nominal model ($R^2$ = 0.83), and 0.86 for the optimized model ($R^2$ = 0.77), using the standard deviation of Yobs as scalar in each case. Thus, the optimized model was declared by C—in stark contrast with the implication by $R^2$—as a big improvement over the nominal model. The improvement, from nominal to optimized model, was in reduction of TotalError from 4.37 to 0.28 units, mostly by a large reduction in SysErr.

To work through the comparison for yourself, first examine the Input&ResultsOverview, YobsYmod&Yp, and ErrorAnalysis&C sheets of HGoFtemplate.xls or ...xlsx, pre-loaded with inputs for assessment of the nominal model for bluegill Wchg; then, copy the appropriate Yobs and Ymod values for WchgBGopt from sheet M=1, ExampleDataSets.xls (or ...xlsx); paste (use "Paste Special," and "Values and numbers format") those into columns D&E of the template's Input&ResultsOverview sheet; notice the immediate change in summary outputs on that sheet; then, examine the YobsYmod&Yp and ErrorAnalysis&C sheets to see other changed results.

*Other Example Applications, and Consequent Conclusions*

We have manipulated other model-vs.-observation datasets, both real and hypothetical, to gain insight and experience as to how C scores for various 1-response systems correspond with the visualized relationships evident in graphical displays. Exploration of those relationships, originating with OrigExmplx1 (Sheet M=1, ExampleDataSets.xls or ...xlsx) is where our search for understanding began; thus, we preserve it here, for the sake of historical fidelity.

Variants of OrigExmplx1 first alerted us to the "Achilles' heel" of our original, standardized-residuals approach (HGoFpres1.avi or ...mov; HGoFpres2.avi or ...mov): OrigExmplx1 has a C-score = 0.93 and $R^2$ = 0.94; two complementary variants, degraded from the original by multiplying each set of modeled outputs either by 2 or by -0.03, yielded C ~ -1.04—but, of course, the same $R^2$, 0.94. The datasets are OrigExmplx1, OrigExmplx2 and OrigExmplx-0.03, from Sheet M=1, ExampleDataSets.xls (or ...xlsx).

From there, we continued to explore various manipulations of OrigExmplx1 (using as scalar the standard deviation of Yobs) and reached the following conclusions:

1) Perfect fit does, in fact, yield C = 1 (dataset OrigExmplYmod=Yobs, from Sheet M=1, ExampleDataSets.xls (and ...xlsx). This particular result holds regardless of the choice in scalar, provided the scalar is the same for all components of error and for all pairs.

2) Fit of a model producing a set of Ymod values identical to the Yobs set, but randomly paired (or a random paring of Ymod and Yobs ranks-without-ties)—which we call the RandMix null model—yields

$E(MSE_{sys}) = N/N = 1;$

$E(MSE_{ran}) = (N-2)/N;$

$E(MSE_{tot}) = 2(N-1)/N;$ and,

$E(C) = 1/N;$ thus,

$E(C) = 0$ in the $N \rightarrow \infty$ limit,

where, E(X) is the expected value of X (see OrigExmpl~PureNoise, from Sheet M=1, ExampleDataSets.xls or ...xlsx).

HGoF_RandMix5.xls (and ...xlsx) is provided by way of evidence in support of the expected-value claims made above. This modification of the generic template enables an all-possible pairing of 5 Ymod values with the identical set of 5 Yobs values (5! = 120 combinations, all with identical means and variances, i.e., those of the 5 Yobs values), followed by computation of respective means of the error terms and of C—which we assert must be their expected values. We already had done the less tedious calculation for the N = 2, 3, and 4 cases, by which we had arrived by intuition and induction at the formulae presented above. For comparison with the N = 5 case, see HGoF_RandMix4.xls (or ...xlsx), for the N = 4 case. Exhaustive random sampling for Yobs datasets involving N from 2 to 500 (CnullRandMixAutoSample1000xMxN.xls or ,,.xlsx) has only reinforced our original conclusion.

We now suggest that the above pattern regarding expected values holds—but with an offset in slope—for the more general situation in which the values of N Ymod are drawn randomly from a quasi-normal distribution (the normal inverse argument being restricted to values between 0.001 and 0.999, the limits corresponding with µ ± 3 σ, approximately) with the same mean and variance as the N Yobs sample.  Like the "RandMix" null model, "RandNorm" presents a "shot-gun blast" of Ymod,Yobs points centered approximately on the Ymod = Yobs line, at the point aveYmod = aveYobs; but, the relation of aveCobs is not 1-to-1 with 1/N; instead, it seems to be 1-to-1 with 1/(2*N):

$$1/(2*N) = - 0.0055 + 1.0074*aveCobs$$

($R^2$ = 0.97, for 48 {aveCobs, 1/(2*N)} sets, each set representing 1,000 random samplings from the RandNorm null, with N ranging from 2 to 500); see summary of these and other results to be described below, at [CnullRandNormAutoSample1000xMxN.xls](#) (or [...xlsx](#)).

Full consideration of these results suggested that, under the RandNorm null,

$\quad$ E(MSEsys) = (N+1)/N;

$\quad$ E(MSEran) = (N-2)/N;

$\quad$ E(MSEtot) = 2 - 1/N = 2*(N-0.5)/N; and,

$\quad$ E(C) = 1/(2*N); thus, again,

$\quad$ E(C) = 0 in the N→∞ limit.

We leave it to others to confirm/reject our conclusions for the generalized relation, by conducting more exhaustive sampling of the RandNorm null.  To facilitate that process, and to enable testing for specific Yobs sets, we provide the template [HGoFtemplateRandNormAutoSample1000.xls](#)  (and [...xlsx](#)). [For sake of comparison, we make available also [HGoFtemplateRandMixAutoSample1000.xls](#)  (and [...xlsx](#)).]

In the sections below, we describe our further exploration of the C vs. N relationship under the RandNorm null, with expansion to accommodate systems involving M > 1, as well.

3) A "Mean Fit" model producing the mean of observed values for each and every observed value—i.e., Ymod = AVERAGE(Yobs), for all N pairs—yields C = (N+1)/2N, with MSEtot = MSEsys = (N-1)/N, and MSEerr = 0.  Thus, C ranges from 1.0 for N = 1, downward to 0.5 as N → ∞.  As random error is introduced (Ymod = AVERAGE(Yobs) + error), C tends to decline.

Consider these two explicit examples: OrigExmplYmod=aveYobs+noErr and OrigExmplYmod=aveYobs+~stdevErr (sheet M=1, ExampleDataSets.xls (or ...xlsx), for both of which N = 10.  For the ...noErr example, C = 0.550 with MSEtot = 0.900.  As Err approaches the standard deviation of Yobs, sampling from the RandNorm null yields average C approaching 1/2N = 0.05 (i.e., average MSEtot approaching 2 - 1/N = 1.9).

Perhaps, more needs be made of the "Mean Fit" circumstance.  First, we acknowledge that successful modeling of a system's average and dispersion of responses, without recourse to the observed data from that system, is no mean feat.

But, when a system is modeled as primarily dependent on its own state (i.e., auto-correlated in time, space or in some other sense), that amounts to a mean fit to extant data, perhaps with added variation that may or may not have a mechanistic basis.  Such models are known in meteorology as "persistence" or "red noise" models (Panofsky and Brier 1968; also, see the U.S. National Weather Service's "Red Noise" webpage):  Tomorrow's weather most likely will be like today's, plus deviations that can be anticipated on the basis of understood or presumed mechanisms.

Suppose Ymod is the mean of some previously observed series, aveYobsPrior, plus a random deviate based on the standard deviation of that previously observed series.  Logic and our testing suggest that C for such a modeled system can be expected to lie between that of the Mean Fit and the Mean Fit + stdevErr scenarios—if the observed system is persistent over the interval of observation and subsequent modeling.  For systems with N = 10, that would mean expected C between ~0.05 and 0.55.  We leave it to those who model (and those who critique models) to consider this issue, when asking, "How good is the model?"

4) A model producing a perfect inverse fit to the data with aveYmod = aveYobs conforms to the equation Ymod = -Yobs + 2*aveYobs.  The perfect-inverse circumstance yields C = -(N-2)/N, with MSEtot = MSEsys = 4(N-1)/N (and, of course, MSEran = 0).   Thus, for N = 10, C = -0.8, with MSEtot = MSEsys = -3.6 and, again, MSEran = 0 (OrigExmpl No-NoiseInverse, from sheet M=1, ExampleDataSets.xls (or ...xlsx).

5) A model producing values representing large positive *or* (not *and*) negative departures from those observed (Ymod = Yobs plus *or* minus a large positive error), yields C << 0, probably → -∞, with no limit to "badness" (OrigExmplWayHigh, and OrigExmplWayLow, from sheet M=1, ExampleDataSets.xls (or ...xlsx).

A sampling of additional one-response systems, representing a variety of model-vs.-observation relationships, is presented in our Excel library, ExampleDataSets.xls (or ...xlsx), sheet M=1 (where M is the number of response variables, 1 for all the examples on this sheet).

*C vs. Conventional Statistical Inference for M = 1 Systems*

We conclude our treatment of the single-response system in simulation modeling, by considering the issue of C's statistical behavior. We accept that increasing C corresponds with improving GoF (defined as declining MSEtot). But, given a model yielding a particular value of C, how do we know whether it is "significant"? That is, can we be confident that a C of such magnitude could arise by chance only with some acceptably low probability? In the absence of extensive information about the sampling distribution of C, perhaps the best that can be done using conventional statistical approaches, is to ask, "Is Ymod under the model yielding a particular value of C, statistically distinguishable from the perfect model, Ymod = Yobs?"

In an attempt to answer this last question, we tried two statistical approaches—one parametric and one non-parametric—for estimating the probability of type-I error for the *deviation* in Ymod from Yobs. These are presented and implemented by the ResidRegrY and WilcoxonY sheets of the generic template [HGoFtemplate.xls](#) (or [...xlsx](#)). In general, C increased with Pr(type-I error) returned by both these conventional statistical analyses—i.e., models with greater values of C also tended to be less statistically resolvable from the data. But, not always.

By way of a first example, we return to our WchgBGnom vs. WchgBGopt comparison (Fig. 1). Application of ResidRegrY does, in fact, declare WchgBGnom to be much more statistically distinct from the perfect relationship Ymod = Yobs than is WchgBGopt: $Pr(F>F') \sim 0.002$ for WchgBGnom vs. $Pr(F>F') \sim 0.884$ for WchgBGopt. But, WilcoxonY yields $Pr(W>W') > 0.2$ for both models, albeit with a greater value of the Wilcoxon statistic W, 27, for WchgBGnom, than the W = 19 for WchgBGopt. (To us, WilcoxonY seems quite reluctant to declare W significant if the linear regression of Ymod on Yobs crosses the Ymod = Yobs line.)

In probing the issue of conventional statistical assessment of C, we encountered another surprise: The one of two models declared statistically more resolvable from the data (more "significant" = lesser probability of type-I error) is not necessarily the model with the lesser C and, thus, the "poorer fit" to those data!

Consider the systems OrigExmpl(Ymod=LoNoiseYobs)x1.05 vs. OrigExmpl(Ymod=HiNoiseYobs)x1.00 (sheet M=1, [ExampleDataSets.xls](#) (or [...xlsx](#)): With regard to modeled vs. observed responses for some system of interest, the target has to be perfect agreement, or Ymod = Yobs. So, logically, regression of Ymod on Yobs would yield a higher value of Pr(type-I error, under the Ymod=Yobs null), for the better-fitting of two models. But, such logic fails...when systematic effect ("signal") and random error ("noise") get in one another's way—and, in the way of common sense.

Look at the two graphs in Figure 3, and decide which model better simulates (or predicts) the observed truth.

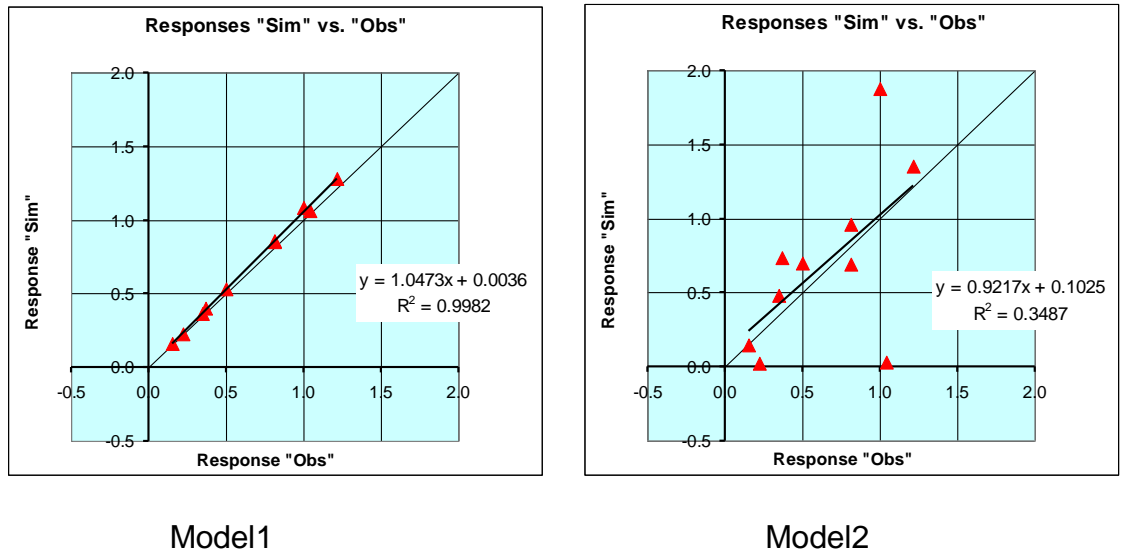

Model1                                                                 Model2

*Figure 3. Simulated responses under two alternative models, Model1 and Model2, versus those observed from a hypothetical system.*

Model1 is Ymod = 1.05*(Yobs, with a small "random" error). Model2 is Ymod = 1.00*(Yobs, with exactly the same series of error values, but with those errors amplified by a factor of 30).

We expect that you will choose Model1. Even though Model1 presents an evident systematic error (with simulated results tending to the high side of those observed—and, increasingly so, as observed values increase), that error tends to be relatively small over the range of the data. Model1's $R^2$, after all, closely approaches 1.0, being 0.998.

On the other hand, Model2 predicts values all over the place, even though its regression line is scarcely further from the 1:1 than is Model1's. And, Model2's $R^2$ is only 0.349.

Thus, Ymod under Model1 is cleanly (and, thus, clearly) resolvable from Yobs; but, Ymod under Model2 is quite "noisy." And, we all know that "noise" is antithetical to good statistical decision-making.

Our Consilience score agrees with your intuition (or, at least, with ours): C is 0.994 for Model1, but only 0.274 for Model2. Thus, it declares Model1 the better model.

In contrast, classical statistics declares Model1 to be way more likely *different* from the perfect model (Ymod=Yobs) than is Model2. Linear regression of Ymod-Yobs residuals on Yobs (with intercept forced to 0); and, Wilcoxon's signed-rank test of pairs (essentially, a non-parametric test of Ho: the median of differences over all pairs is 0) both declare Model2 to be far *less* distinguishable from the perfect model (Y=X) than is Model1. For Model1, probabilities under the null hypothesis, Ymod=Yobs, are vanishingly small; but, for Model2 those probabilities exceed 0.2:

| Pr(type-I error, Ho:Ymod=Yobs) under | Model1 | Model2 |
|---|---|---|
| Linear regression of Y-X residuals on X, with intercept forced to 0 | 9.3E-05 | 0.839 |
| Wilcoxon's signed-rank test for pairs | <0.001 (W=0) | >0.2 (W=19) |

Conclusion: <u>Statistically harder-to-reject (larger Pr(type-I error, under Ymod=Yobs null)) does not mean better-fitting, in the sense of minimizing error between model and observation</u>.

On the other hand, patently poor models (e.g., Ymod = 5 + noise, where the maximum of Yobs is only 1.22) also have small Pr(Ymod=Yobs), under the same statistical tests—but their values of C decline, to and through 0, and on toward -∞. This is consistent with the truism that a model can be no better than perfect—Ymod=Yobs, → C=1—but there is no finite limit to how "bad" (wrong) a model can be.

**C vs. GoF Measures of Schunn and Wallach (2005)**

Among the comparative GoF measures reviewed by [Schunn and Wallach (2005)](#), we judge their RMSSD—actually, its antecedent square, which we will call MSSD—to be the most closely related to our C.

We interpret MSSD to be the mean of $((Yobs_i - Ymod_i)/(stdevYobs_i/\sqrt{n_i}))^2$. I.e., this approach assumes that each pair's Yobs, $Yobs_i$, comes with its own standard-error estimate. Because the N $stdevYobs_i$ likely differ from one another, this means MSEtot for the dataset cannot be partitioned into only the two components MSEsys and MSEran, without regard for the possibility of a non-zero mean cross-product.

Our computed C for the [Schunn and Wallach's (2005)](#) example dataset (N = 12; Sheet M=1, [ExampleDataSets.xls](#) or [...xlsx](#)) is 0.690, not far removed from the dataset's $R^2$, 0.684—such agreement between C and $R^2$ to be expected, given that the regression of Ymod on Yobs lies close to the one-to-one line, Ymod = Yobs (with MSEsys accounting only for 22.3 % of MSEtot).

For this dataset, [Schunn and Wallach (2005)](#) report that RMSSD is 2.90 (we calculate 2.85, the difference presumably the result of our rounding errors); thus, MMSD is $(2.90)^2$ = 8.41. How good or bad is the fit indicated by an MMSD = 8.41? A larger value would imply poorer fit. But there are no convenient "landmarks" that would guide the inexperienced interpreter of RMMSD or MMSD scores.

Nor is there an evident way for evaluating the overall performance of a model that simulates or otherwise generates a separate MMSD score for each of M types of response to be compared with those from an observed system.

**Joint C: Weighting of Multiple Responses for HGoF**

In the general circumstance, where M > 1, concern must shift to how the separate C values might be integrated, to yield a single joint C.

We approached the problem by seeking a weighting scheme such that each of M component responses receives weight in proportion to its degree of independence from the other M -1 components, with the sum of weights required to be 1. Covariance of Yobs series was deemed the most reasonable basis for such weighting.

Example: Suppose M = 3, with the component response variables named Y1, Y2, and Y3. If Y1obs and Y2obs are perfectly correlated (either positively or negatively), and Y3obs is scarcely correlated with either Y1obs or Y2obs, then we really have only 2 response variables—1) Y1&Y2, and 2) Y3; and, the $R^2$ half-matrix looks like

|       | Y1obs | Y2obs | Y3obs |
|-------|-------|-------|-------|
| Y1obs | 1     | 1     | ~0    |
| Y2obs |       | 1     | ~0    |
| Y3obs |       |       | 1     |

Weights logically should be 0.25 for Y1, 0.25 for Y2, and 0.5 for Y3.

Now, instead, suppose the $R^2$ half-matrix looks like

|       | Y1obs | Y2obs | Y3obs |
|-------|-------|-------|-------|
| Y1obs | 1     | 0.5   | 0.5   |
| Y2obs |       | 1     | 0.5   |
| Y3obs |       |       | 1     |

This means all 3 (*2) possible pairs of the 3 response variables have equal $R^2$ (and, in fact, as far as weighting is concerned, it is irrelevant what that $R^2$ actually is). So, the appropriate weight is 0.333 for each.

Here is the weighting formula we worked out, for computing the weight for the ith of M component responses (M > 2*):

> Weight for the $i^{th}$ of M component responses:
>
> $$W_i = \frac{1}{M}\left[1 + \frac{(M-2)}{(M-1)}\left[\frac{\sum R^2_{i-excl}}{(M-2)} - \frac{\sum R^2_{i-incl}}{2}\right]\right]$$
>
> where $R^2_{i-excl}$ is $R^2$ for observed-response pair excluding $i^{th}$ component response, and $R^2_{i-incl}$ is $R^2$ for observed-response pair including $i^{th}$ component response.
>
> E.g., $M = 5, i = 3$:
>
> $$R^2_{i-excl} = \{R^2_{1,2}, R^2_{1,4}, R^2_{1,5}, R^2_{2,4}, R^2_{2,5}, R^2_{4,5}\}$$
>
> $$R^2_{i-incl} = \{R^2_{1,3}, R^2_{2,3}, R^2_{3,4}, R^2_{3,5}\}$$

*For M = 2, the only logical values of Wi are 0.5, 0.5; thus, the two component C's are simply averaged to compute joint C. And for M = 1, W must be 1.

SUM($R^2$ i-excl) and SUM($R^2$ i-incl) are, respectively, the sums of $R^2$ values for (M-1)*(M-2)/2 pairs of components excluding (i-excl) and the (M-1) pairs including (i-incl) the ith component. (Both sets exclude cases of self-correlation, from the diagonal of the half-matrix.)

Note that ours is an empirical weighting function, at which we arrived largely by "reverse-engineering." We have run many test datasets through the function, and it has not failed yet, in a logical sense. Check it yourself, for the M = 3 situations outlined above, or for your own test data, to see if you get sensible results. The HGoF template implements our co-variance weighting scheme automatically, for M up to 5.

Two additional considerations: First, our scheme—which automatically weights for relative sample size—does not preclude further weighting, for other relevant dimensions, e.g., economic and/or socio-political importance. It is only necessary that the final set of M weights add to 1.0. Given Wi, relative sample-size effNi, and relative-importance value RIi, the final weight for Ci becomes Wi*effNi*RIi/SUMall(W*effN*RI). The HGoFtemplate enables easy implementation of such further weighting, for RI, or "Anything Else" (AE)—see B3:B7, "WiCalc" sheet.

Second, we considered those situations in which the Ni cases (records) producing Yiobs bear no particular relation to the Nj cases producing Yjobs: CaseMatch? = *No.*

Suppose, for example, that our "ShrimpTaFeedFinalModPatchy25" dataset (sheet M=5, ExampleDataSets.xls or ...xlsx—sorry for the need to get ahead of ourselves here; see below re "patchy" datasets in general and this dataset in particular) had arisen *not* as a set of responses linked by case (albeit with some missing values)—CaseMatch? = *Yes*; but rather, that the M responses had been observed for different groups of individual shrimp, in different experiments, perhaps even in different labs, etc.; i.e., CaseMatch? = *No*. Another way of expressing the distinction is to suppose that there had been *no* overlap between cases observed for two or more of the M responses. Nevertheless, a single model, invoking the same ruleset, had produced a legitimate value of Ymod for each Yobs across all M and Ni.

We decided that HGoFtemplate could and should accommodate CaseMatch? = *No* situations. Accordingly, the present version of HGoFtemplate sets CORREL(Yiobs,Yjobs)$^2$ = 1/(MIN(Ni,Nj) - 1)—which we claim is the expected value for random association (see RSQ_RandMix5.xls or ...xlsx; and, RSQtemplateRandMixAutoSample1000.xlsx)—if CaseMatch? = *No*; and, to the computed value of CORREL(Yiobs,Yjobs)$^2$ [= RSQ(Yiobs,Yjobs)] if CaseMatch? = *Yes*. In addition, for CaseMatch? = *No*, effNij is set equal to MIN(Ni,Nj), instead of being set equal to the number of overlaps between Yiobs and Yjobs.

On the "Input&ResultsOverview" sheet of HGoFtemplate, the user can set CaseMatch? for each Yi,Yj combination to *Yes* or *No*, in U3:X6. The default value is *Yes*. And, for all datasets in ExampleDataSets.xls (or ...xlsx), CaseMatch? is universally *Yes*.

To explore CaseMatch?, paste the "ShrimpTaFeedFinalModPatchy25" dataset into HGoFtemplate, with M set to 5. Confirm that default summary results are as presented for that dataset, on sheet M=5 of ExampleDataSets.xls (or ...xlsx). Now, change all CaseMatch? values to *No* and note the changed values of effN, from 8.1 to 10.2; and, of joint C, from 0.792 to 0.838. For this particular example, we interpret the changes as reflecting what essentially amounts to increased effective sample size, and the greater evenness in values of Wi, causing Yi with higher C to have increased representation.

Situations involving non-uniform AE and/or CaseMatch? = *No* are accommodated also by CompactCCalculator.xlsx. The user need only reset the relevant cells within the Calculator's blocks T3:T7 and U28:AA31. Again, the default conditions are uniform AE (= 1) and CaseMatch? = *Yes*, universally.

It is important to realize that both these "considerations" have to do only with *weighting*. They have no bearing on the values of C for individual response pairs Yiobs,Yimod.

*Example Applications, M = 3 and 5:  Shrimp Ecophysiology and Growth*

(We skip over the M = 2 circumstance, because application involves only simple averaging to arrive at joint C.  But, 3 two-response datasets relating to simulation of fish growth in weight vs. length—WLmod1, WLmod2, and WLmod3—are included on sheet M=2, ExampleDataSets.xls (or ...xlsx); and, for a Camtasia presentation describing the three simulation models and analysis of associated datasets, access 1) HGoFpres1.avi or ...mov; and, 2) HGoFpres2.avi or ...mov.)

We already have referenced two M = 1 datasets associated with application of the simulation model Ecophys.Fish (Neill et al. 2004) to the bluegill, a freshwater sunfish.  Co-author Walker (2009; see also Walker et al. 2009, 2011) did extensive experiments on ecophysiology and growth of a marine shrimp, then adapted and applied an elaboration of Ecophys.Fish, named Ecophys.Shrimp, to aid in interpreting the results of his experiments.  Here, we present applications of HGoF to three Ymod-vs.-Yobs datasets arising from Walker's (2009) research.  For each of these three datasets, there were at least 3 response variables (M = 3)—Wchg (growth rate), RMR (routine metabolic rate), and MMS (marginal metabolic scope).

ShrimpTaFeedEarlyMod75 and ShrimpTaFeedFinalMod75 (Sheet M=3, ExampleDataSets.xls (or ...xlsx) focus on the same random sample of 75 experiments (of 102 total) with individual shrimp subjected to temperature and feeding treatments; in each of these experiments, Wchg, RMR, and MMS were among the responses measured.

Observed results were compared with those simulated under an early version and a final version of Walker's model, Ecophys.Shrimp.  Covariance-based joint C for ShrimpTaFeedEarlyMod75 was only 0.318; for ShrimpTaFeedFinalMod75, joint C was markedly higher, 0.797.  This comparison suggests that the final model is a substantial improvement over the early model.

Would the same final model provide similarly good fit in simulation of representative results from the full diversity of Walker's shrimp experiments—experiments that included the added treatments of salinity, dissolved-oxygen acclimation, and initial shrimp size?  To find out, we evaluated Ymod vs. Yobs for a random sample of 500 experiments (of 644 total), named "ShrimpAllExperFinalMod500"—and determined that covariance-based joint C differed only slightly from that computed with the 75-experiment subset:  0.775 for ShrimpAllExperFinalMod500 (N=500), vs. 0.797 for ShrimpTaFeedFinalMod75 (N=75).

Focusing more closely on ShrimpAllExperFinalMod500, covariance-based weight for Wchg was about 0.46, vs. just over 0.27 for each of the two other responses, RMR and MMS, both of which logically and mechanistically are more closely related to metabolism than is biomass growth rate.  Despite the weighting asymmetry, covariance-based joint C scarcely differed from joint C based on equal weighting, because the separate contributions of the three component responses to joint C were so similar—0.809 for Wchg, 0.791 for RMR, and 0.703 for MMS.

For the 102 experiments randomly sampled to yield the 75-experiment dataset ShrimpTaFeedFinalMod75, Walker (2009) also measured, and modeled as integral components of Ecophys.Shrimp, two additional response variables, %Surv (percentage of the treatment group surviving) and LOCr (dissolved-oxygen concentration limiting for routine metabolism). The aggregate dataset (sheet M=5, ExampleDataSets.xls or ...xlsx) provides an example for HGoF analysis in the M = 5 circumstance. Note that the values for Wchg, RMR, and MMS are the same here as for the equivalent M = 3 dataset.

Covariance-based joint C for the M = 5 analysis of ShrimpTaFeedFinalMod75 was 0.807, about the same as the 0.797 for the M = 3 case involving the same individual shrimp. The range of component C's was 0.645 (%Surv) to 0.978 (LOCr).

**Application to Multiple Regression and Analytical Models**

It occurred to us, in afterthought, that the same HGoF analysis might be applied to multiple regression (MR) and analytical models. To explore that idea, we went to the Internet for a 2-response multiple-regression example and developed our own example dataset to demonstrate application to a set of three related 1-response analytical models.

Our MR example is taken from a homework exercise assigned by Dr. D.W. Stockburger in his graduate multivariate statistics course at Missouri State University. We don't know the expected solutions to the exercise, but we used Excel to arrive at the regression analysis shown in MultRegrExampleMoSU.xls and ...xlsx, with the resultant dataset MultRegrExample (sheet M=2, ExampleDataSets.xls or ...xlsx). HGoF analysis of this dataset yields joint C = 0.819, with components 0.970 for Y1 and 0.668 for Y2.

In contrast with applications of HGoF to simulation models, for least-squares MR models, the means of all 4 error components (systematic, non-systematic, interaction, and total) always will be zero. For simulation models, only the non-systematic ("noise") and interaction mean errors always have a zero mean.

Analytical models are like simulation models (and unlike MR models) in that they regularly present a non-zero mean for systematic error. The example we developed has Y1 being actual area of a 2-dimensional "shape" versus its estimated area calculated as that of a circle with the same visualized mean diameter (N = 10 shapes); Y2 being ranked actual area of shape versus ranked area based on that calculated as before; and, Y3 being ranked actual area of shape versus ranked visualized "size," the latter with sorting and ordering permitted. The system and its analytical modeling are schematized in Figure 4 and made available for closer examination and manipulation in Shapes.ppt.

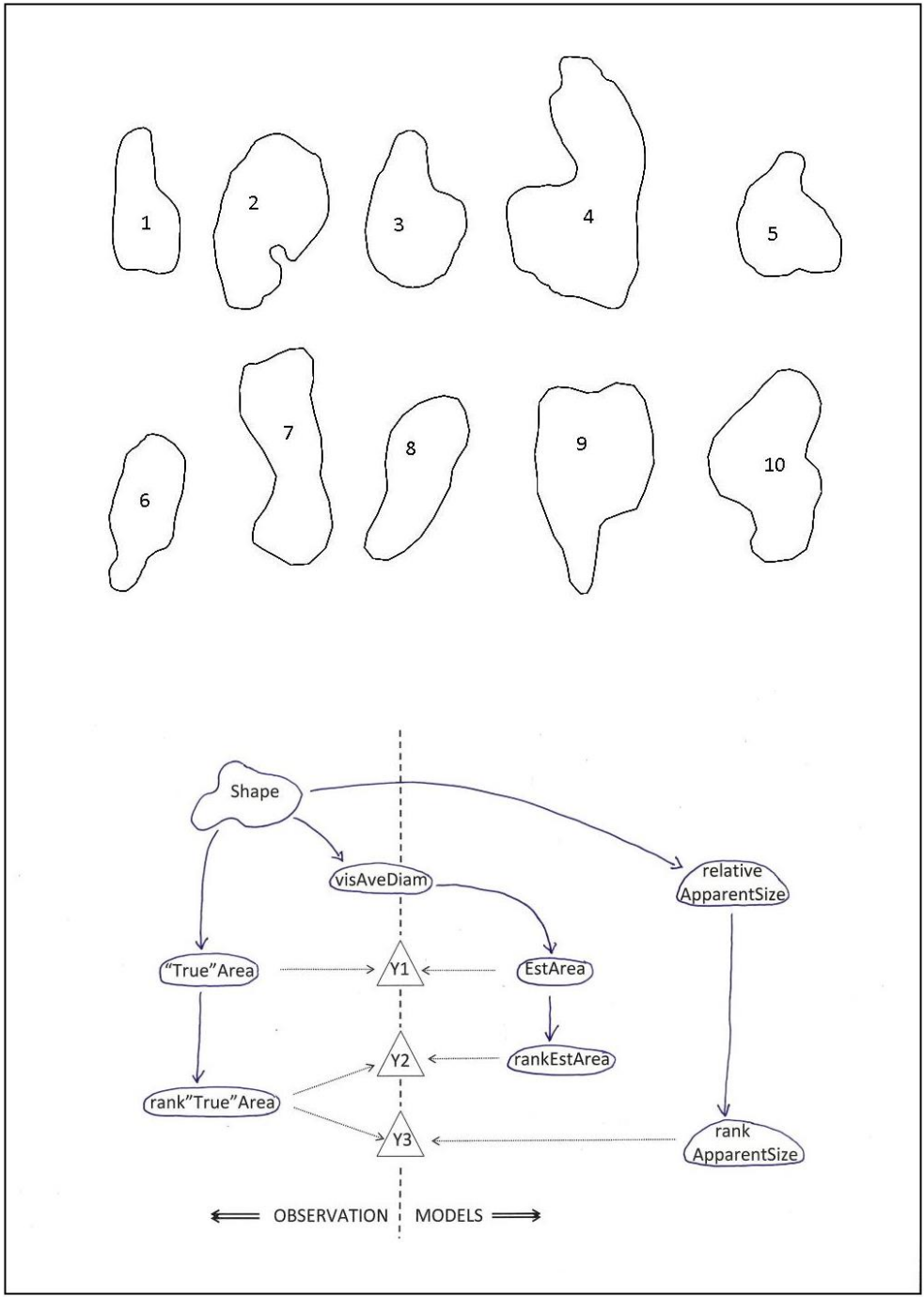

*Figure 4. Analyical modeling of the absolute and relative apparent sizes of 2-dimensional shapes. Above are the ten sample shapes; below is the conceptual*

*relationship between observed and modeled systems. "True"Area of each shape is that measured by application of SketchAndCalc[TM]; rank"True"Area is the rank of Shape by its "True"Area, from smallest to largest. EstArea is calculated as π\*(visAveDiam/2)$^2$, where visAveDiam is that declared upon the senior author's visual inspection of shape, then measurement of the resulting straight line's length. The ordinal rankApparentSize is the rank obtained by ordering shapes from smallest to largest according to relativeApparentSize, as visualized by the senior author. Tied ranks are assigned their average values.*

Results are presented as AnalyticalModel.xls and ...xlsx. In this instance, it turns out that visualization of relative apparent size—sort of a "Gestalt" approach—yields results more consilient (C = 0.967) with "ground truth" than is what might seem to be the more objective approach—i.e., considering the Shape as approximating a circle, for which only the average diameter need be declared, to estimate area (C = 0.757). But, the latter approach does have the advantage of not needing to depend on visual access to more than one of the 10 shapes at a time.

But, back to the first sentence of the paragraph-before-last: "Analytical models are like simulation models (and unlike MR models) in that they regularly present a non-zero mean for systematic error." However, note that the mean systematic error will be zero if the Ymod and Yobs are presented as ranks-without-ties (as in this example)—or, if the Ymod are restricted to be any 1-to-1 rearrangement of the Yobs set. Further, in the case of ranks-against-ranks, the perfect-inverse relationship (as poor as fit can be, in the case of ranks-without-ties) always will present C = -(N-2)/N, or -0.8 in the case of N = 10 pairs.

**Probabilistic Assessment of C and Joint C**

Toward the end of the section "*Other Example Applications, and Consequent Conclusions*," we alluded to an assessment of C's distribution under the null model Ymod = ~Normal(aveYobs, stdevYobs), as a function of the number of pairs, N. By simply ranking the 1000 outcomes for each set by C, we arrived in each case at an estimate of C's probability distribution under the null model. Further, by examining the null distributions of C and joint C for several additional datasets—with M > 1—we reached this conclusion: The relevant measure of "sample size" is M\*effN, which reduces to N for M = 1.

We propose that critical values of C and joint C are described effectively by a series of empirical reverse power-hyperbolic functions of X = LOG10(M\*effN),

$$C'(\alpha) = 1 - (X^n / (X_{0.5}^n + X^n))\ ,$$

where C'(α) = the critical value of C at Pr(C>C')~α. M is, as before, the number of component responses. For M = 1, effN = N; for M > 1, effN is the average number of Yobs,Ymod pairs in overlap (see below, for more explanation regarding effN). Finally, n and $X_{0.5}$ are parameters estimated for each value of α.

Below are trial-and-error estimates of the parameters n and $X_{0.5}$; also presented, for each pair of parameter estimates, are consequent values of $R^2$ for the least-squares regression of C' modeled vs. observed for all 69 datasets:

| α | n | $X_{0.5}$ | $R^2$ |
|---|---|---|---|
| 0.01 | 2.85 | 25.0 | 0.99 |
| 0.05 | 2.50 | 15.0 | 0.99 |
| 0.10 | 2.25 | 11.0 | 0.99 |
| 0.25 | 1.90 | 4.5 | 0.97 |
| 0.50 | 1.70 | 2.3 | 0.93 |

Joint C for this set of five models was 0.968 (N = 69; "CnullRandNormAutoSample1000xMxN_29Sep17.xls, BestModC's," sheet M=5, ExampleDataSets.xls or ...xlsx).

Figure 5 provides a graphical presentation of the family of empirical functions for alpha from 0.01 to 0.5, using the parameters tabled above.

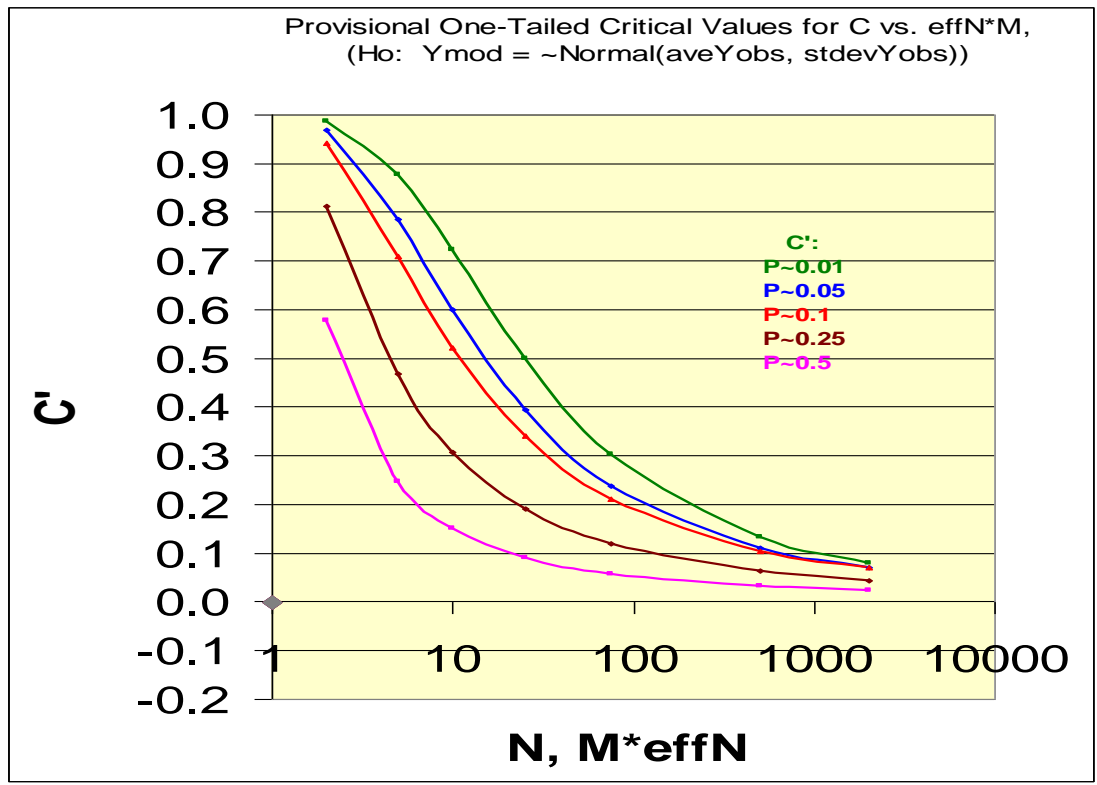

*Figure 5. Critical C vs. M\*effN at 5 levels of alpha, the probability of type-I error, under the null model Ymod = ~Normal(aveYobs, stdevYobs). This graph is consistent with the nomogram on the JointC sheet of* HGoFtemplate.xls *(and* ...xlsx*).*

Note that our treatment of critical C makes no formulaic distinction between C for a given Yi, and joint C for M Yi's. Thus, critical C at alpha = 0.05 is the same for a single Yi with N pairs = 30, as for joint C in an M = 3, all N =10, system (—it is ~ 0.36).

Although critical C's were functions of M\*effN, average value of joint C—thus, its expectation—was a function only of effN. Adding 21 additional analyses for "joint C," to the 48 for "single C," scarcely altered the regression of 1/(2\*effN) on aveCobs:

$$1/(2*effN) = -0.0049 + 1.0088*ave(Cobs, or jointCobs)$$

($R^2$ = 0.975, for 69 {ave(Cobs, or jointCobs), 1/(2\*N)} sets, each set representing 1,000 random samplings from the RandNorm null, with N ranging from 2 to 500, and M from 1 to 5); again, see the summary of all these results, at CnullRandNormSampleMxNx1000.xls (or ...xlsx).

Systems involving differences in N, and/or patchiness, among the M response variables required extra consideration, in that the product M\*N itself logically should be adjusted for variation in N. What we decided, was to substitute for M\*N the product M\*effN, where effN is the system average of effNi,j, the number of overlapping Yobs,Ymod pairs for each relevant combination Yi,Yj. Thus,

$$effN = [SUM(effN_{i,j}) \text{ for all possible } i,j <=M]/[M*(M-1)/2].$$

As an example of this computation, we use "ShrimpTaFeedFinalModPatchy25," Sheet M=5, ExampleDataSets.xls or ...xlsx, with

N1=25,   N2= 10,   N3= 8,   N4=20,   N5=10;

where, the values of effNi,j (from inspection of cells AF13:AO13, sheet "Input&ResultsOverview," HGoFtemplate.xls (or ...xlsx) are

|    | Y1 | Y2 | Y3 | Y4 | Y5 |
|----|----|----|----|----|----|
| Y1 | *  | 10 | 8  | 20 | 10 |
| Y2 |    | *  | 2  | 7  | 4  |
| Y3 |    |    | *  | 7  | 5  |
| Y4 |    |    |    | *  | 8  |
| Y5 |    |    |    |    | *  |

The effNij sum for all Yi,Yj combinations is 81, with the average = 81/(M*(M-1)/2) = 81/10 = 8.1 = effN.

If M is restricted to 4, effN is 54/6 = 9.0; for M = 3, effN = 20/3 = 6.67; and for M = 2, effN = 10.  Note that this N-weighting scheme returns the constant N if all M responses have one and the same N, regardless of M.

HGoFtemplate.xls and ...xlsx incorporate, on their "JointC" sheets, a nomogram that auto-plots the computed C values for up to M = 5 Yi's, and for their joint C, on a graph of C'(α) vs. M*effN, with isopleths for α = 0.5, 0.25, 0.1, 0.05, and 0.01.  All supporting data are provided in and through CnullRandNormSampleMxNx1000.xls (and in ...xlsx), which contains hyperlinks to the 25 Excel files that fully document the C' analysis.

We end this section by restating a point made above:  A model that reliably delivers values of C and joint C in the neighborhood of 0, and therefore "not significant," relative to the null model, Ymod = ~Normal(aveYobs, stdevYobs), still may reflect important truth and be useful.  This is to say that modeling the mean and variation of Yobs is non-trivial, if estimation of the parameters aveYobs and stdevYobs is independent of the subject Yobs series.

**Conclusion**

Our HGoF analysis and its emergent statistic "Consilience" (C) offer a holistic approach for quantifying goodness-of-fit between outputs modeled and observed for some system of interest.  HGoF is complementary to conventional statistical analysis, which assesses probabilistic effects of independent variables on potentially dependent responses. HGoF asks to what degree does a network of presumed cause-and-effect—whether inferred via statistical analysis or not—produce outcomes in agreement with those observed.  Moreover, HGoF affords a partition of lack-of-fit into systematic and non-systematic components, thus facilitating the revision of hypotheses that comprise the subject model.

The statistical behavior of C is made tractable by its explicit basis in error analysis.  C is a function only of the mean of squared scaled-error between paired modeled and observed values (Ymod and Yobs), MSEtot:

$$C = -(MSEtot - 2)/2 \ .$$

We have been able to identify some "landmarks" for C, when computed with the standard deviation of Yobs as scalar; note that all are relatively simple functions of N, the number of paired modeled and observed results:

|  | MSEsys | MSEran | MSEtot | C |
|---|---|---|---|---|
| Perfect Fit: Ymod = Yobs, pairing preserved | 0 | 0 | 0 | 1 |
| Mean Fit: Ymod = AveYobs | (N-1)/N | 0 | (N-1)/N | (N+1)/2N |
| Mean Fit with "Noise": |  |  |  |  |
| Ymod = Yobs, but randomly paired | N/N = 1 | (N-2)/N | 2(N-1)/N | 1/N |
| Ymod = random sample from quasi-normal distribution with mean = aveYobs and SD = stdevYobs | (N+1)/N | (N-2)/N | 2(N-0.5)/N | 1/(2N) |
| Perfect Inverse Fit: Ymod = -Yobs + 2*AveYobs | 4(N-1)/N | 0 | 4(N-1)/N | -(N-2)/N |

These measures are for the component responses of systems observed and modeled. For systems with three or more component responses of interest (M >= 3), we have proposed integrating the component C values by weighting them in proportion to the relative independence of the M observed datasets (Y1obs, Y2obs,...YMobs) from one another, to compute a joint C. The weight for the $i^{th}$ of M response components is

$W_i = (1/M)*[1 + ((M-2)/(M-1))*(\Sigma R^2_{i\text{-excl}}/(M-2) - \Sigma R^2_{i\text{-incl}}/2)]$ , IFF M >= 3; ELSE 1 for M = 1, 0.5 for M = 2.

For Yobs series declared logically independent of one another (i.e., Yiobs and Yjobs arising from unrelated cases), $R^2$ is set to what we have estimated is its expected value for random association, 1/(N-1); or, to 1/(MIN(Ni,Nj)-1), in the event Ni and Nj differ (see below). Otherwise, $R^2$ is set to its computed nominal value, RSQ(Yiobs,Yjobs).

Our analysis accommodates datasets for which N is not constant over M, and for which there are vacant Yobs,Ymod pairs within series ("patchy" datasets). In such cases, a measure of effective N, reflecting the variation in N and in the average overlap of Yi,Yj series, is computed for the entire dataset, and returned as effN. For fully "balanced" datasets, with N constant over M, and complete overlap between series, effN = N. For unbalanced datasets, additional weighting is performed in accordance with the response series' contributions to effN.

Provision also is made for further weighting of component responses, to accommodate any other factor declared relevant by the user.

Our assessment of C and joint-C distributions, computed with the standard deviation of Yobs as scalar and with co-variance-based weighting, suggests that critical values under the null hypothesis Ymod = ~Normal(aveYobs, stdevYobs) can be approximated as empirical functions of the product M*effN. Our HGoF template offers a nomogram enabling probabilistic evaluation of computed C and joint-C values under the null model.

It is our hope that colleagues will apply, evaluate and extend our HGoF approach, using their own ideas and data. We urge that issues of autocorrelation and other aspects of distribution among the N elements within series, and interactions among the M series, be considered and probed. We acknowledge that our conclusions regarding distributions of C and joint C, and their relation to N and M, amount to empirical conjectures arrived at mainly by brute force. The need for a stronger, more mathematically-secure basis is apparent.

We offer the associated Excel templates, together with the CompactCCalculator, to make the application and testing of HGoF easy. We would ask that users let us know of successes and problems. We are especially hopeful for informative responses from those studying weather dynamics and climate change, ecosystem process and structure, and issues of human health and well-being.

**Hyperlinks Cited**

AnalyticalModel.xls = http://people.tamu.edu/~w-neill/Consilience/AnalyticalModel27Sep16.xls
...xlsx = http://people.tamu.edu/~w-neill/Consilience/AnalyticalModel27Sep16.xlsx

Appendix I = http://people.tamu.edu/~w-neill/Consilience/Investigating the error cross-product term.docx

CnullRandMixAutoSample1000xMxN.xls = http://people.tamu.edu/~w-neill/Consilience/CnullRandMixAutoSample1000xMxN_29Sep17.xls
...xlsx = http://people.tamu.edu/~w-neill/Consilience/CnullRandMixAutoSample1000xMxN_29Sep17.xlsx

CnullRandNormAutoSample1000xMxN.xls = http://people.tamu.edu/~w-neill/Consilience/CnullRandNormAutoSample1000xMxN_29Sep17.xls
    ...xlsx = http://people.tamu.edu/~w-neill/Consilience/CnullRandNormAutoSample1000xMxN_29Sep17.xlsx

CompactCCalculator.xlsx = http://people.tamu.edu/~w-neill/Consilience/CompactCCalculatorCXv3.xlsx [xls version unavailable]

EcoFish = http://people.tamu.edu/~w-neill/WFSC417617/EcoFishPresShort600x800.html

Ecophys.Fish = http://people.tamu.edu/~w-neill/EcophysFish/EcophysFishHomepage.htm

ExampleDataSets.xls = http://people.tamu.edu/~w-neill/Consilience/ExampleDataSets29Sep17.xls
    ...xlsx = http://people.tamu.edu/~w-neill/Consilience/ExampleDataSets29Sep17.xlsx

Fisher's exact-probability test = https://en.wikipedia.org/wiki/The_Design_of_Experiments

Handbook of Biological Statistics = http://www.biostathandbook.com/

HGoFpres1.avi = http://people.tamu.edu/~w-neill/Consilience/HGoFpres1WAS17Feb09.avi
    ...mov = http://people.tamu.edu/~w-neill/Consilience/HGoFpres1WAS17Feb09.mov

HGoFpres2.avi = http://people.tamu.edu/~w-neill/Consilience/HGoFpres2Sequel20Sep14.avi
    ...mov = http://people.tamu.edu/~w-neill/Consilience/HGoFpres2Sequel20Sep14.mov

HGoF_RandMix4.xls = http://people.tamu.edu/~w-neill/Consilience/HGoF_RandMix4-2Feb16.xls
    ...xlsx = http://people.tamu.edu/~w-neill/Consilience/HGoF_RandMix4-2Feb16.xlsx

HGoF_RandMix5.xls = http://people.tamu.edu/~w-neill/Consilience/HGoF_RandMix5-16Oct16.xls
    ...xlsx = http://people.tamu.edu/~w-neill/Consilience/HGoF_RandMix5-16Oct16.xlsx

HGoFtemplate.xls = http://people.tamu.edu/~w-neill/Consilience/HGoFtemplateM5-18Mar18.xls

    ...xlsx    = http://people.tamu.edu/~w-neill/Consilience/HGoFtemplateM5-18Mar18.xlsx

HGoFtemplateRandMixAutoSample1000.xls  =  http://people.tamu.edu/~w-neill/Consilience/HGoFtemplateRandMixAutoSample1000-M1to5_31Aug17.xls
    ...xlsx  =  http://people.tamu.edu/~w-neill/Consilience/HGoFtemplateRandMixAutoSample1000-M1to5_31Aug17.xlsx

HGoFtemplateRandNormAutoSample1000.xls  =  http://people.tamu.edu/~w-neill/Consilience/HGoFtemplateRandNormAutoSample1000-M1to5_14Aug17.xls
    ...xlsx  =  http://people.tamu.edu/~w-neill/Consilience/HGoFtemplateRandNormAutoSample1000-M1to5_14Aug17.xlsx

MR example = http://www.psychstat.missouristate.edu/multibook/mlt06.htm

MultRegrExampleMoSU.xls     = http://people.tamu.edu/~w-neill/Consilience/MultRegrExampleMoSU.xls
    ...xlsx    = http://people.tamu.edu/~w-neill/Consilience/MultRegrExampleMoSU.xlsx

Pearson's chi-square = https://en.wikipedia.org/wiki/Pearson%27s_chi-squared_test

RSQ_RandMix5.xls = http://people.tamu.edu/~w-neill/Consilience/RSQ_RandMix5.xls
    ...xlsx = http://people.tamu.edu/~w-neill/Consilience/RSQ_RandMix5.xlsx

RSQtemplateRandMixAutoSample1000.xlsx = http://people.tamu.edu/~w-neill/Consilience/RSQ_RandMixN.xlsx
[Execute with Excel 2007 or later; xls version unavailable.]

Schunn and Wallach (2005)  = http://www.lrdc.pitt.edu/schunn/gof/Schunn&Wallach-GOF.pdf

Shapes.ppt = http://people.tamu.edu/~w-neill/Consilience/Shapes.ppt

SketchAndCalc[TM] = https://www.SketchAndCalc.com

U.S. National Weather Service's "Red Noise" webpage  = http://www.nws.noaa.gov/om/csd/pds/PCU2/statistics/Stats/part2/Noise_red.htm

**Appendix I**.   Investigating the error cross-product term.docx

Also available from the people.tamu.edu FTP server are down-loadable copies of this ms., as

http://people.tamu.edu/~w-neill/Consilience/ConsilienceMs.21Oct18.pdf .

**NOTE:  All people.tamu.edu-hyperlinked files cited in the body of this document and listed above in "Hyperlinks Cited"—together with all files hyperlinked within those files—are available for download individually, as a sub-set within one of 3 subfolders, or as the entire set, from WHN's Mega-account folder "Consilience and Holistic Goodness-of-Fit (HGoF)," via https://mega.nz/#F!wUVHVajD!8Fv85WqLLoSjyxZZ0_ZwSQ .  (Sorry, but following-the-link *may* require that highlighted URL be copied-and-pasted or typed into browser address window.)**

**This back-up measure has been necessitated by increasingly problematic and unreliable performance of the people.tamu.edu server.**